\documentclass[pdflatex, sn-mathphys-num, oneside]{sn-jnl}

\usepackage{graphicx}%
\usepackage{multirow}%
\usepackage{amsmath,amssymb,amsfonts}%
\usepackage{amsthm}%
\usepackage[title]{appendix}%
\usepackage{xcolor}%
\usepackage{textcomp}%
\usepackage{manyfoot}%
\usepackage{booktabs}%
\usepackage{algorithm}%
\usepackage{algorithmicx}%
\usepackage{algpseudocode}%
\usepackage{listings}%
\usepackage{braket}
\usepackage[version=4]{mhchem} 

\newcommand{\etacoll}{(97 \pm 2)\,\%} 

\newcommand{\etaprod}{(84.7 \pm 10.6)\,\%} 

\newcommand{\qeff}{(87.3 \pm 11.1)\,\%} 

\newcommand{\gammamax}{117.5\,\mathrm{\mu s}^{-1}} 

\newcommand{\fluxbroad}{(99.5 \pm 12.5)\,\mathrm{Mcps}}

\newcommand{\fluxzpl}{(24.0\pm 2.1)\,\mathrm{Mcps}}

\newcommand{\indistinguishabilty}{(91.2 \pm 2.3)\,\%}

\theoremstyle{thmstyleone}%
\theoremstyle{thmstyletwo}%

\theoremstyle{thmstylethree}%

\begin{document}

\title[Article Title]{
100 million photons per second from a single organic molecule
}



\author[1,2]{\fnm{Siwei} \sur{Luo}} 
\author[1,2]{\fnm{Tim} \sur{Hebenstreit}} 
\author[1]{\fnm{Alexey} \sur{Shkarin}}
\author[1]{\fnm{Jan} \sur{Renger}}
\author[1]{\fnm{Tobias} \sur{Utikal}}
\author*[1,2,3]{\fnm{Stephan} \sur{Götzinger}}\email{stephan.goetzinger@mpl.mpg.de}

\affil[1]{\orgname{Nano-Optics Division, Max Planck Institute for the Science of Light}, \orgaddress{\postcode{91058} \city{Erlangen}, \country{Germany}}}

\affil[2]{\orgdiv{Department of Physics}, \orgname{Friedrich Alexander University Erlangen-Nuremberg}, \orgaddress{\postcode{91058} \city{Erlangen}, \country{Germany}}}

\affil[3]{\orgdiv{Graduate School in Advanced Optical Technologies (SAOT)}, \orgname{Friedrich Alexander University Erlangen-Nuremberg}, \orgaddress{\postcode{91058} \city{Erlangen}, \country{Germany}}}



\abstract{
At cryogenic temperatures, organic single-photon sources (SPSs) offer
Fourier-limited emission, virtually unlimited photostability, and emission
wavelengths selectable by molecular design, making them attractive for quantum
metrology, secure communication, and photonic quantum information processing.
However, their efficient integration with photonic microstructures has remained
challenging, limiting photon collection from single organic emitters under
cryogenic operation. Here, we demonstrate a cryogenic planar metallo-dielectric
antenna that efficiently directs the emission of a single dibenzoterrylene (DBT)
molecule towards the collection optics. The device reaches a collection
efficiency of $97\,\%$ and delivers $10^{8}$~photons per second into the first
lens. We measure a photon indistinguishability for the Fourier-limited
transition of $91.2\,\%$ while maintaining a high single-photon purity of
$97.7\,\%$ under strong continuous-wave~(CW) and pulsed excitation. These
results establish organic molecules as a high-performance platform for
single-photon generation and represent an important step towards scalable
organic quantum photonic technologies.
}



\maketitle

\section{Introduction} \label{introduction}

Photonic quantum technologies, from quantum communication~\cite{gisin2007quantum,couteau2023applications} to quantum networks and optical quantum computing~\cite{couteau2023applications,o2007optical}, rely on the on-demand generation of single photons. This need has driven intense efforts to develop robust, scalable and high-performance SPSs based on a wide range of quantum emitters, including single atoms~\cite{kuhn2002deterministic,volz2006observation,wilk2007single,ritter2012elementary}, trapped ions~\cite{keller2004continuous,blinov2004observation}, semiconductor quantum dots~\cite{lounis2000photon,michler2000quantum,shields2007semiconductor}, and colour centres in diamond~\cite{kurtsiefer2000stable,togan2010quantum,aharonovich2014diamond} and silicon~\cite{redjem2020single,higginbottom2022optical}. Several of these platforms now combine near-unity single-photon purity, high photon indistinguishability and advanced pho-
tonic integration, with semiconductor quantum dots reaching a particularly high level
of technological maturity \cite{maring2024versatile,loredo2026deterministic}.

Organic molecules offer a qualitatively distinct paradigm for single-photon generation \cite{shkarin2026organic}.
As chemically defined emitters with atomically precise, reproducible structures,
their transition energies, oscillator strengths, and chemical functionality can be engineered by synthesis, enabling emission across a broad spectral range from the visible to the near-infrared. Molecular emitters can moreover be designed to host addressable spin degrees of freedom~\cite{roggors2026single}, adding a further functionality to the platform. Their nanometer-scale dimensions allow high emitter densities and compact integration, while low fabrication costs and compatibility with scalable processing make them attractive
for hybrid and large-area photonic architectures~\cite{huang2025chip}.
Different molecular species can be co-doped into a common crystalline matrix,
providing a chemically programmable route to spectrally multiplexed quantum emitters
within a single material system. Together, these properties position organic molecules as a quantum light source platform whose degrees of freedom are unlocked by molecular design rather than by nanofabrication.
 
Single molecules in crystalline solid-state matrices provide excellent single-emitter optical coherence. The most widely studied systems are polycyclic aromatic hydrocarbons (PAHs) embedded as guest molecules in organic host crystals~\cite{toninelli2021single,adhikari2022progress}. The host supplies a rigid environment that suppresses dephasing and can enable Fourier-limited optical transitions~\cite{trebbia2010indistinguishable,rezai2018coherence,lombardi2021triggered}. Although local variations lead to inhomogeneous broadening, optimized molecular guest--host systems show ensemble distributions on the scale of a hundred gigahertz \cite{nicolet2007single}. Here we use DBT embedded in a 1,4-dichlorobenzene (\ce{\textit{p}-DCB}) crystal, as illustrated in Fig.~\ref{fig_introduction}a. At temperatures below $2$~K, this system combines Fourier-limited optical transitions with effectively unlimited photostability~\cite{verhart2016spectroscopy}, as required for high-performance single-photon generation.

More broadly, photon extraction is a fundamental limitation for solid-state SPSs: photons not captured by the first optical element are irretrievably lost and cannot be recovered by improved detectors, lower-loss optics or more efficient fibre coupling. First-lens collection efficiency is therefore a stringent and comparatively setup-independent measure of source performance. In conventional bulk or thin-film geometries, quantum emitters radiate into a broad angular distribution, while refractive-index mismatch traps a large fraction of the emission by total internal reflection, a problem common to all solid-state emitters regardless of host index~\cite{barnes2002solid}. This universal extraction problem has motivated a range of photonic engineering approaches, including microcavities~\cite{li,somaschi2016near,wang2019towards,barbiero2022high,mao2025single} and nanowire geometries~\cite{claudon2010highly,babinec2010diamond}.

For cryogenic molecular SPSs, however, these solutions have remained largely out of reach: the crystalline organic matrices required for high optical coherence are fragile and chemically incompatible with the lithographic and etching processes used in micro- and nanofabrication. The main obstacle to state-of-the-art performance of organic emitters has therefore not been their intrinsic properties, but the efficiency of photon extraction.

A particularly suitable route to overcoming this bottleneck is provided by planar metallo-dielectric antennas. We previously introduced such structures as broadband, non-resonant antennas capable of reaching collection efficiencies above $99\,\%$ at room temperature~\cite{chu2014experimental}. The concept relies on a layered geometry in which the emitter is placed in a dielectric layer between media of different refractive indices~\cite{chu2014experimental,chen2018highly,chu2017single}. This quasi-waveguide directs emission preferentially towards the high-index side, while a metallic mirror redirects light that would otherwise be lost. Since the antenna mechanism relies on geometric light redirection combined with waveguide-mode leakage rather than a narrow optical resonance, it is broadband and largely insensitive to dipole orientation. These properties make the antenna particularly well suited for cryogenic molecular emitters: efficient extraction is achieved without any micro- or nanostructuring of the crystalline host itself.

Here we demonstrate a cryogenic planar metallo-dielectric antenna for a single organic molecule and show that molecular SPSs can reach collection efficiencies and brightness levels surpassing leading solid-state platforms. The device combines planar films in a nanochannel with a directly bonded solid immersion lens and is operated with single DBT molecules embedded in \ce{\textit{p}-DCB}. Using single-molecule spectroscopy and Fourier-plane imaging, we determine a first-lens collection efficiency of approximately $97\,\%$ into a lens of numerical aperture $\mathrm{NA}=0.77$. Within the $744$--$842\,\mathrm{nm}$ detection band, which contains the majority of the molecular emission, we measure a detected count rate of $21.3\,\mathrm{Mcps}$, an order-of-magnitude increase over the detected rates of previous cryogenic single-molecule sources \cite{colautti20203d,lombardi2020molecule,lombardi2021triggered,kuck2022single}. Accounting for the detection efficiency and the spectral distribution of the
emission, this corresponds to approximately $10^{8}$~photons per second
collected into the first lens, to our knowledge the highest first-lens photon
collection rate reported for a cryogenic SPS, independent of emitter platform. Spectral filtering of the $0$--$0$ zero-phonon line yields a detected count rate of $8.9\,\mathrm{Mcps}$, corresponding to $24.0\,\mathrm{Mcps}$ collected into the first lens. The source maintains near-unity single-photon purity, with $g^{(2)}(0)$ remaining below $2.5\,\%$ even under strong continuous-wave and pulsed excitation, and shows a photon indistinguishability of $\indistinguishabilty$, limited by residual pure dephasing. These results establish antenna-integrated organic molecules as exceptionally bright and optically coherent SPSs whose emission wavelength is set by molecular design, opening a route towards scalable organic quantum photonic devices.

\section{Results} \label{results}

\subsection{Design and implementation of the metallo-dielectric antenna} \label{subsec_antenna_design}

\begin{figure}[!htbp]
    \centering
    \includegraphics{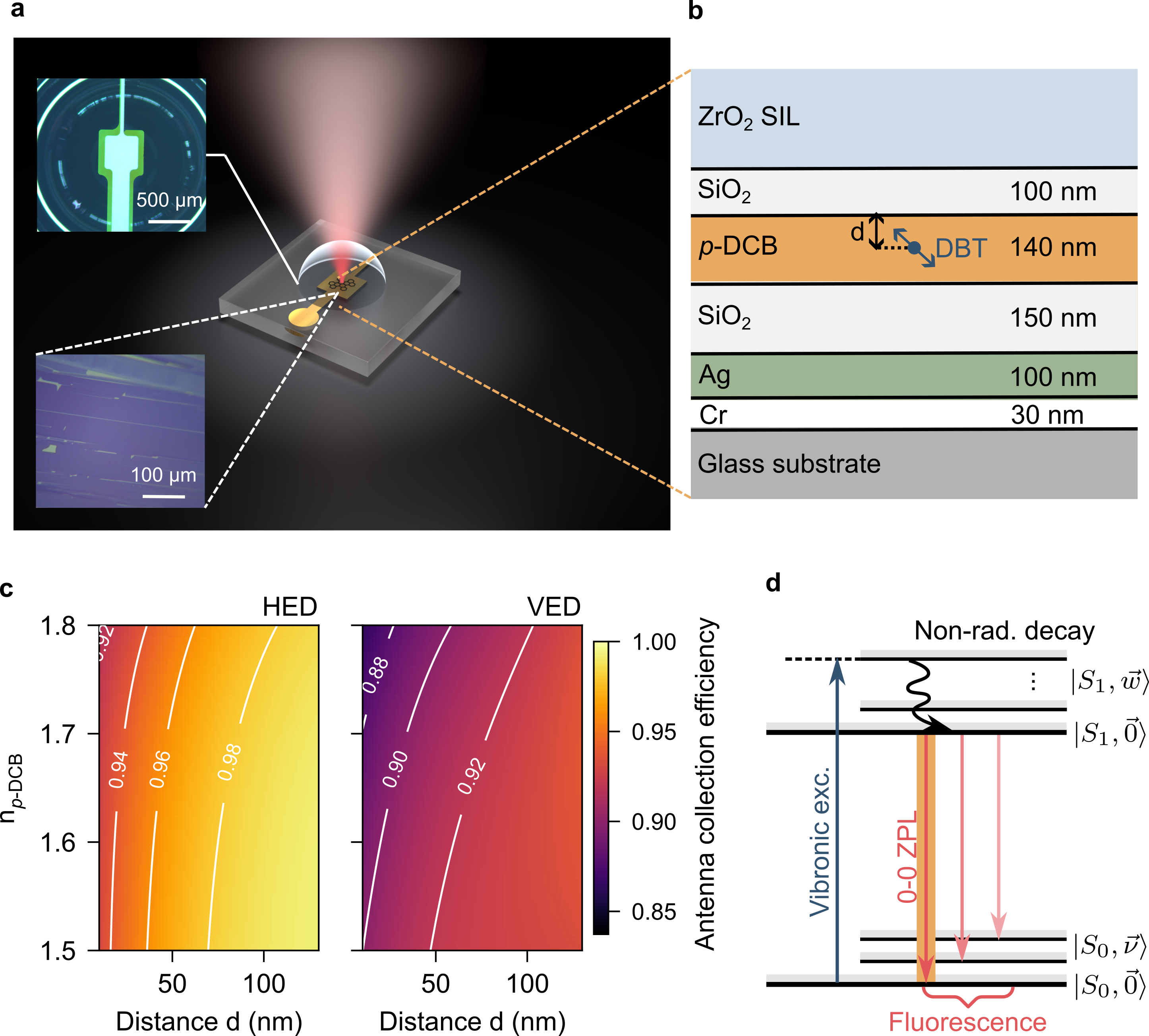}
    \caption{\textbf{Illustration of the metallo-dielectric antenna incorporating organic materials.} \textbf{a} Schematic of the  antenna structure. Upper inset: Microscope image of the antenna viewed through the flat substrate prior to crystal filling. Lower inset: Cross‑polarized microscope image of \ce{\textit{p}-DCB} inside the channel. The crystal grows preferentially along a specific direction (horizontal in the plot), forming an array of elongated crystalline stripes. \textbf{b} Antenna layer structure with detailed thicknesses. \textbf{c} Simulated collection efficiency of the antenna using a $0.77\,\text{NA}$ aspheric lens, presented as a function of the refractive index of \textit{p}-DCB, dipole position $d$ within the nanochannel, and dipole orientation. HED: horizontal electric dipole; VED: vertical electric dipole. \textbf{d} Energy level diagram of a DBT molecule, along with excitation and detection schemes for single-molecule spectroscopy. $S_0$ and $S_1$ denote the ground and excited electronic manifolds, respectively, with their vibronic sublevels labeled by $\vec{\nu}$ and $\vec{w}$.}
    \label{fig_introduction}
\end{figure}

Planar antenna concepts for efficient photon collection have been extensively explored at room temperature in the past \cite{lee2011planar,chu2014experimental,chu2017single}. However, these implementations typically rely on high-NA oil-immersion objectives and are therefore not compatible with cryogenic operation. To meet the requirements of low-temperature experiments, we employ a hemispherical \ce{ZrO2} solid immersion lens (SIL, $\mathrm{n}=2.14$) as the high-index collection medium coated with a $100\,\text{nm}$ layer of \ce{SiO2}, as shown in Fig.~\ref{fig_introduction}a. The rest of the antenna is realized using a patterned substrate in which a silver mirror and a protective \ce{SiO2} spacer are deposited, resulting in a channel thickness of $140\,\text{nm}$, as sketched in Fig.~\ref{fig_introduction}b. The SIL is then directly bonded to this substrate, forming a monolithic structure, as confirmed by the absence of interference fringes at the interface, shown in the upper inset of Fig.~\ref{fig_introduction}a. The resulting nanochannel is subsequently filled with molten \ce{\textit{p}-DCB}, which solidifies into a homogeneous organic layer hosting the DBT molecules. As illustrated in the lower inset of Fig.~\ref{fig_introduction}a, cross-polarized optical microscopy reveals that the organic layer consists of crystal domains extending over several hundred micrometers, so that the material behaves as an effectively homogeneous medium with minimal scattering. A comprehensive treatment of the antenna design optimization and fabrication process is given in Ref.~\cite{luo2026designing}. 

To assess the performance of the antenna, we carried out analytical calculations following the formalism of \citeauthor{chen2007efficient} \cite{chen2007efficient,chen201199}. In these calculations, the birefringence of \ce{\textit{p}-DCB} is captured by treating it as a homogeneous medium with a tunable refractive index between 1.5 and 1.8, covering the accessible range of the effective refractive index of the quasi-waveguide mode. We further account for uncertainties in emitter position and dipole orientation. The resulting collection-efficiency map is shown in Fig.~\ref{fig_introduction}c.  Overall, the antenna exhibits a high degree of robustness against both dipole displacement and orientation, maintaining collection efficiencies above $90\,\%$ over a wide parameter range. A moderate reduction is observed when the emitter approaches the SIL interface, which can be attributed to enhanced evanescent coupling and
corresponding leakage at angles beyond the acceptance range of the collection optics~\cite{luo2026designing}.

\subsection{Cryogenic single-molecule spectroscopy and optical excitation scheme} \label{subsec_spectroscopy}

We characterize individual molecules embedded in the antenna using cryogenic single-molecule spectroscopy under both CW and pulsed confocal excitation. The relevant energy-level scheme is shown in Fig.~\ref{fig_introduction}d, where two electronic manifolds are considered, each accompanied by a set of vibronic sublevels. In our experiments, the molecule is excited to a higher vibronic state and subsequently relaxes through non-radiative channels to the vibrational ground state of the first excited electronic manifold, $\ket{S_{1},\vec{0}}$.

For CW excitation, the laser is tuned to the strongest vibronic transition, located $291\,\mathrm{cm}^{-1}$ above $\ket{S_{1},\vec{0}}$ \cite{zirkelbach2022high}, in order to maximize the excitation efficiency. Radiative decay from $\ket{S_{1},\vec{0}}$ to vibrational levels $\ket{S_{0},\vec{\nu}}$ of the electronic ground state then produces a molecular fluorescence spectrum composed of several zero-phonon lines (ZPLs) and their associated phonon sidebands. This fluorescence is intrinsically broadband.

Excitation via a vibrationally excited state offers two key advantages over direct resonant excitation of the spectrally narrow $0$-$0$ ZPL. First, it allows the system to reach population inversion, which efficiently populates the excited state. Second, it keeps the excitation and emission wavelengths spectrally well separated. This separation facilitates detection of the brightest and narrowest transition, namely the $0$-$0$ ZPL. To isolate this line, we employ a tunable notch filter based on a volume Bragg grating. The resulting narrowband single-photon stream, highlighted in yellow in Fig.~\ref{fig_introduction}d, is then routed to the detection system for measurements of brightness, single-photon purity, and indistinguishability.

\subsection{Detection of fluorescence intensity} \label{subsec_intensity}

\begin{figure}[!htbp]
    \centering
    \includegraphics{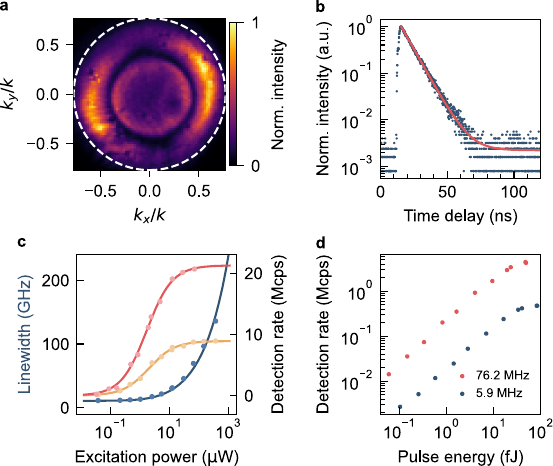}
    \caption{\textbf{Characterization of collection efficiency and fluorescence intensities.} \textbf{a} Back focal plane image of $0$-$0$ ZPL photons from a single DBT molecule. The dashed white circle marks the acceptance angle of the aspheric lens ($\mathrm{NA} = 0.77$). \textbf{b} Time-resolved fluorescence decay (blue) and its single-exponential fit (red) for a DBT molecule inside the antenna. \textbf{c} Power-dependent excitation linewidth (blue) and fluorescence detection rate for the $0$-$0$ ZPL (yellow) and broadband emission (red) under CW vibronic excitation. \textbf{d} Power-dependent fluorescence detection rate of the $0$-$0$ ZPL under pulsed non-resonant excitation at repetition rates of $76.2 \, \text{MHz}$ (red) and $5.9 \, \text{MHz}$ (blue).}
    \label{fig_intensity}
\end{figure}

We next evaluate the fluorescence collection enabled by the antenna.
Fig.~\ref{fig_intensity}a shows the angular emission pattern of the 0--0 ZPL of
a single DBT molecule, imaged in the Fourier plane using a Bertrand lens.
Two features in the image are relevant for the analysis of antenna collection
efficiency. First, the signal drops to the background level at the edge of the
accessible NA (dashed white circle). Since the $140$~nm thick \textit{p}-DCB
layer supports no quasi-waveguide mode beyond the collection
NA~\cite{luo2026designing}, all angular emission maxima are confined within the
accessible NA, and the emission decreases smoothly from the maximum towards
larger angles. The absence of signal at the rim thus implies that no light is
lost above the critical angle. Consequently, the molecule cannot sit close to
the SiO$_2$ layer adjacent to the SIL, where it would emit a substantial
fraction of its light into these large angles. Second, the pattern exhibits a
pronounced azimuthal modulation, as expected for a dipole with a predominantly
in-plane component~\cite{luo2026designing}. Evaluating the calculation of
Fig.~\ref{fig_introduction}c under these two constraints yields a collection
efficiency of $\eta_\text{CE}=\etacoll$ into the first lens.

We then quantify the source brightness by CW and pulsed power-dependent measurements. The fluorescence decay trace of the selected molecule is shown in Fig.~\ref{fig_intensity}b. A single-exponential fit yields an excited-state lifetime of $\tau=(8.51 \pm 0.02)\,\text{ns}$, corresponding to a maximum emission rate of $(117.5 \pm 0.3)\,\text{Mcps}$ under vibronic excitation. In this analysis, intersystem crossing is assumed to be negligible, consistent with the bunching statistics in the intensity autocorrelation function.

Under CW vibronic excitation, we record a series of excitation scans centred at
$291\,\mathrm{cm}^{-1}$ for increasing excitation powers. After background
subtraction using a spatial excitation scan via a fast steering mirror, each
spectrum is fitted with a Lorentzian profile.
The extracted linewidths and fluorescence count rates are shown in
Fig.~\ref{fig_intensity}c. To determine the total brightness of the source, we
detect in this measurement not only the 0--0 ZPL but also most of the red-shifted emission up to a short-pass filter edge at $842\,\mathrm{nm}$. With increasing
power, the detected fluorescence saturates at $(21.3 \pm 0.3)\,\text{Mcps}$.
Correcting for the measured optical transmission, detector efficiency of the
setup and for the residual emission beyond $842\,\mathrm{nm}$, this corresponds to a photon flux of $\fluxbroad$ into the first
lens over the full fluorescence spectrum. The uncertainty is dominated by that
of the Franck--Condon factor, $\alpha = (33 \pm 3)\,\%$, which depends on the
phase of the host crystal~\cite{zirkelbach2022high}. For all single-photon
measurements below, only the 0--0 ZPL is used; isolating it, we detect
$(8.9 \pm 0.1)\,\text{Mcps}$, corresponding to $\fluxzpl$ collected into the
first lens.

At full saturation, the excitation probability approaches unity, so that the saturated broadband flux corresponds to a fraction $\etaprod$ of the maximum emission rate $\Gamma = 1/\tau = \gammamax$. This fraction is the product of the quantum efficiency of the molecule and the collection efficiency of the antenna. Since both quantities are bounded by unity, this fraction is at the same time a lower bound on each of them. This value is therefore an experimental bound on the quantum yield of the DBT molecule investigated. Combining it with the collection efficiency derived above yields a quantum efficiency of $\qeff$.

To operate the selected emitter as a triggered, on-demand SPS, we also characterize the emission of the 0--0 ZPL under pulsed excitation. We use repetition rates of $76.2\,\text{MHz}$ and $5.9\,\text{MHz}$, with the latter obtained by additional pulse picking such that the emitter fully relaxes to its ground state between successive pulses, satisfying the condition for clean, deterministic single-photon emission. With increasing pulse energy, the fluorescence saturates (Fig.~\ref{fig_intensity}d). The maximum detected count rates are $4.6\,\text{Mcps}$ and $480.0\,\text{kcps}$ at the two repetition rates, corresponding to 0-0 ZPL collection rates at the first lens of $(12.5\pm1.1)\,\text{Mcps}$ and $(1.3\pm0.1)\,\text{Mcps}$, respectively. From these measurements, we extract the per-pulse collection probability, a key metric for triggered sources that quantifies their ability to deliver photons deterministically on demand. Considering the full emission spectrum, this probability reaches $90.8^{+9.2}_{-11.5}\,\%$, which is slightly higher than the quantum efficiency obtained in the CW case---an effect we attribute to non-ideal leakage pulses between consecutive main pulses, which can also excite the molecule. Such collection efficiency ranks this cryogenic emitter among the most efficient organic SPSs ever reported, while simultaneously leveraging the $0$--$0$ ZPL as a low-background source of indistinguishable photons for practical applications.

\subsection{Single-photon purity} \label{subsec_purity}

Single-photon purity is a key figure of merit for any SPS, as it directly reflects the underlying photon statistics. In our device, an optimized crystal-filling strategy strongly suppresses background fluorescence and enables nearly background-free operation of the embedded molecules. We quantify the source purity by measuring the second-order autocorrelation function, $g^{(2)}(\tau)$, using a Hanbury Brown-Twiss (HBT) setup (see Methods).

\begin{figure}[!htbp]
    \centering
    \includegraphics{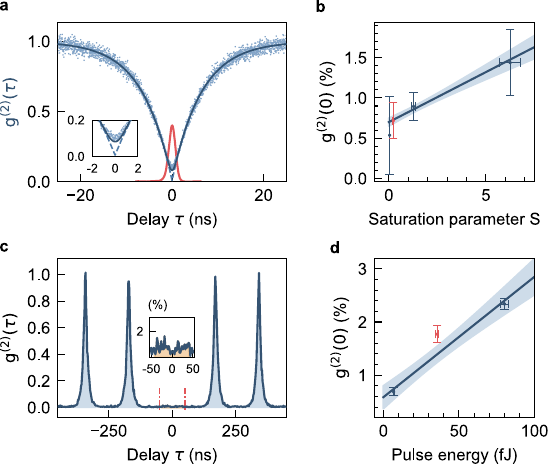}
    \caption{\textbf{Single-photon purity under CW and pulsed excitation.} \textbf{a}, Second-order autocorrelation histogram measured under CW excitation. The experimental data are shown in light blue. The solid and dashed curves represent fits including and excluding detector timing jitter, respectively. The measured instrument response function (IRF) is shown in red. \textbf{b}, Power dependence of $g^{(2)}(0)$ under CW excitation, together with a linear fit (blue). The result shown in \textbf{a} is highlighted in red, and the light-blue shaded area denotes the $95\,\%$ confidence band. \textbf{c}, Second-order autocorrelation histogram measured under pulsed excitation. Inset: enlarged view of the central correlation region between the two red lines. \textbf{d}, Power dependence of $g^{(2)}(0)$ under pulsed excitation, together with a linear fit (blue). The result shown in \textbf{c} is highlighted in red.}
    \label{fig_purity}
\end{figure}

A representative $g^{(2)}(\tau)$ histogram recorded under CW excitation at a saturation parameter of $S = 0.23 \pm 0.02$ is shown in Fig.~\ref{fig_purity}a. The raw antibunching contrast is reduced by the finite detector timing response, as confirmed by the two-detector instrument response function (IRF), which has a linewidth of approximately $2\,\text{ns}$ and is therefore comparable to the emitter lifetime. To account for this effect, we fit the data with a model in which the ideal autocorrelation function is convolved with the measured IRF. This yields $g^{(2)}(0) = (0.7 \pm 0.2)\,\%$, where the uncertainty is obtained from the fit. The autocorrelation functions before and after convolution are shown as dashed and solid lines, respectively, in Fig.~\ref{fig_purity}a. We next perform power-dependent measurements of $g^{(2)}(0)$ under CW excitation, as shown in Fig.~\ref{fig_purity}b. In the low-background regime, $g^{(2)}(0)$ scales linearly with the ratio of background contribution, and the data are correspondingly well described by a linear fit. These measurements indicate high single-photon purities even at a high saturation parameter.

Under pulsed excitation, second-order correlation measurements are performed at a repetition rate of $5.9\,\text{MHz}$, such that the individual correlation peaks are well separated in time. The autocorrelation is evaluated by comparing the peak area around zero delay with the average area of side peaks at large delay, as indicated by the yellow and blue shaded regions in Fig.~\ref{fig_purity}c. Because direct integration can be biased by temporally uncorrelated background, which raises the baseline, we analyze the histogram using the post-processing method introduced in \cite{baltisberger2026indistinguishable}, which explicitly accounts for the baseline level and the finite integration window. This yields a corrected value of $g^{(2)}(0) = (1.8 \pm 0.2)\,\%$. The uncertainty of the central peak is dominated by shot noise, whereas the uncertainty of the side peaks is obtained from the standard deviation of the selected correlation peaks. Power-dependent measurements under pulsed excitation reveal a similar linear increase of $g^{(2)}(0)$, but with a slightly larger slope than in the CW case (Fig.~\ref{fig_purity}d). At the maximum count rate, we obtain $g^{(2)}(0) = (2.3 \pm 0.1)\,\%$. 
We attribute the elevated residual multiphoton contribution to the higher energy demands imposed by the inefficient broadband pulsed excitation. In addition, avalanche photodiode (APD) breakdown flashes \cite{kurtsiefer2001breakdown} induce some optical crosstalk between the detectors and contribute to the residual central peak. This interpretation is supported by the double-peak structure visible in the central correlation region, shown in the inset of Fig.~\ref{fig_purity}c.

\subsection{Photon indistinguishability} \label{subsec_indistinguishability}

Many photonic quantum information protocols rely on two-photon interference and therefore require photons that are identical in all relevant degrees of freedom. The standard test of photon indistinguishability is the Hong-Ou-Mandel (HOM) effect, in which two photons impinging simultaneously on a beam splitter interfere. This quantum interference suppresses coincidence events between the two output ports and thereby provides a direct measure of indistinguishability. It is therefore a key figure of merit, as excess dephasing from residual spectral diffusion or molecular dynamics directly reduces the interference contrast. To probe the indistinguishability of photons from a single molecule, we delay one photon and interfere it with a subsequent one at a beam splitter. Experimentally, this is implemented using a fibre-based Mach--Zehnder interferometer (MZI) incorporating a fibre delay line and a tunable polarization controller, following approaches previously established for molecular emitters \cite{rezai2018coherence,lombardi2021triggered,schofield2022photon}.

\begin{figure}[!htbp]
    \centering
    \includegraphics{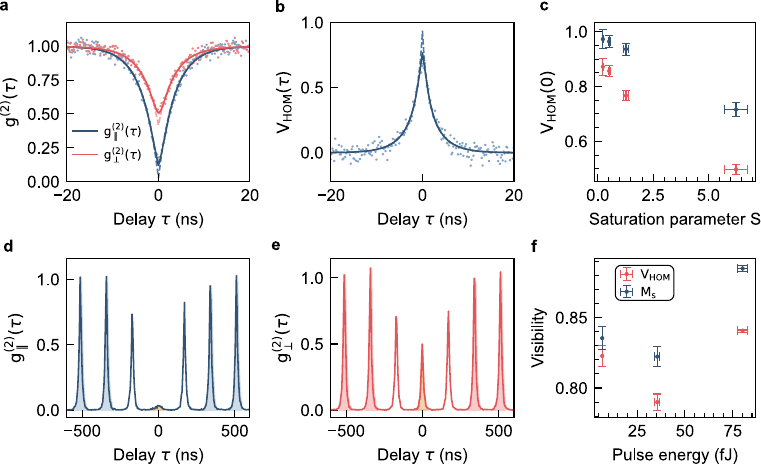}
    \caption{\textbf{Photon indistinguishability under CW and pulsed excitation.} \textbf{a}, Two-photon interference (TPI) measurements under CW excitation for orthogonal and parallel input polarizations. \textbf{b}, HOM visibility extracted from the data in \textbf{a}, shown with and without accounting for the detector response. \textbf{c}, Power dependence of the zero-delay HOM visibility, $V_{\text{HOM}}(0)$, extracted from raw data (red) and after correction for detector timing response (blue). \textbf{d,e}, TPI histograms measured under pulsed excitation for parallel and orthogonal input polarizations, respectively. The central peak area is highlighted in yellow, whereas the side peaks used for normalization are indicated by the corresponding line colours. \textbf{f}, Power dependence of the HOM visibility, $V_{\text{HOM}}$, and the corrected single-photon indistinguishability, $M_{s}$, obtained from peak-area integration.}
    \label{fig_indistinguishabilty}
\end{figure}

Under CW excitation, two-photon interference manifests itself as a suppression of coincidence events around zero delay. For ideally distinguishable and indistinguishable photons, this corresponds to $g^{(2)}(0)=0.5$ and $g^{(2)}(0)=0$ for orthogonal and parallel input polarizations, respectively. The measured correlation traces are shown in Fig.~\ref{fig_indistinguishabilty}a. A pronounced antibunching dip is observed in both cases, while the finite detector timing response reduces the contrast of the central feature. We analyze the data using the theoretical framework of ref.~\cite{lombardi2021triggered}, incorporating the independently measured $g^{(2)}(\tau)$ of the source at the same saturation parameters $S$ used in Fig~\ref{fig_purity}b. The corresponding idealized traces without detector timing jitter are shown as dashed lines and recover the intrinsic molecular response, with the coincidence dips approaching their expected limiting values.

From the two measurements, we extract the HOM visibility,
$
V_{\text{HOM}}(\tau)=\frac{g^{(2)}_{\perp}(\tau)-g^{(2)}_{\parallel}(\tau)}{g^{(2)}_{\perp}(\tau)},
$
following refs.~\cite{keller2004continuous,patel2008postselective}. The resulting visibility traces, with and without accounting for the instrument response function, are shown in Fig.~\ref{fig_indistinguishabilty}b. Evaluated at zero delay, the visibility reaches $(76.7\pm1.8)\,\%$ for the raw data and $(93.6\pm2.2)\,\%$ after correcting for detector timing jitter. Repeating the measurement over a range of saturation parameters yields the data shown in Fig.~\ref{fig_indistinguishabilty}c, where visibilities extracted from raw and deconvoluted data are shown in red and blue, respectively. The uncertainties are obtained by propagation of the fit errors. Below saturation ($S<1$), the corrected visibility exceeds $90\,\%$, demonstrating a high degree of indistinguishability at $\tau=0$. At higher excitation powers, the visibility decreases, which we attribute to photo-induced spectral diffusion acting as an additional dephasing channel.

Under pulsed excitation, the influence of detector timing jitter on the extracted HOM visibility is negligible, provided that the interferometer delay is matched to the pulse separation so that two consecutively emitted photons overlap at the final beam splitter of the MZI. In our case, the delay is $\Delta t = 170.7\,\text{ns}$, highlighting that photons emitted with such a large temporal separation can still interfere with high visibility. The corresponding correlation histograms for parallel and orthogonal polarizations are shown in Figs.~\ref{fig_indistinguishabilty}d,e. The histogram is normalized to the average peak amplitude of the correlation peaks, excluding the three central peaks, which are associated with fewer than three classes of interference events \cite{loredo2016scalable}. The HOM visibility is then obtained from the central peak areas according to
$
V_{\text{HOM}}=\frac{A_{\perp}-A_{\parallel}}{A_{\perp}},
$
where $A_{\perp}$ and $A_{\parallel}$ denote the normalized central peak areas for orthogonal and parallel polarizations, respectively. As for the pulsed $g^{(2)}(0)$ analysis, we determine the baseline and optimize the integration window following the procedure described above. The resulting background-corrected data yield a raw visibility of $V_{\text{HOM}}=(84.1\pm0.1)\,\%$, with the uncertainty obtained by error propagation.

To isolate the indistinguishability of the single-photon component, we first
correct the raw visibility for the residual multi-photon contribution, using the
independently measured single-photon purity from Fig.~\ref{fig_purity}d.
Following ref.~\cite{ollivier2021hong}, the multi-photon component is treated as
separable noise added to an ideal SPS, such that
\[
M_{s}=\frac{V_{\text{HOM}}+g^{(2)}(0)}{1-g^{(2)}(0)}.
\]
This yields $M_{s}=(88.5\pm0.2)\,\%$. This value still includes imperfections of
the HOM setup itself, in particular non-ideal polarisation control and
interferometer imbalance. To separate these from the properties of the emitter,
we analyse the data with the model of ref.~\cite{schofield2022photon}, which
accounts for these setup imperfections together with an excess dephasing rate
$\gamma$. The fit yields $\gamma=(5.7\pm1.6)\,\mathrm{\mu s}^{-1}$, corresponding to an
emitter-limited indistinguishability of
\[
M_{\gamma}=\frac{\Gamma_{1}}{\Gamma_{1}+2\gamma}=(91.2\pm2.3)\,\%,
\]
where $\Gamma_{1}=1/\tau$ denotes the population relaxation rate. The higher
value of $M_\gamma$ reflects the removal of the setup imperfections, which are
included in the model but not in the area-based analysis. Excess dephasing
therefore emerges as the dominant limitation to unity indistinguishability.
Finally, we extend the pulsed indistinguishability measurements to a wider range of pulse energies, as shown
in Fig.~\ref{fig_indistinguishabilty}f. Across the three investigated pulse
energies, we obtain an average indistinguishability $M_s$ of approximately
$85\,\%$, with the residual variation attributed primarily to laser-induced
fluctuations of the excess dephasing.

\section{Discussion} \label{discussion}

The performance of our source rests on resolving a long-standing trade-off in
molecular SPSs: the crystalline hosts that provide the best optical coherence
are precisely those that are incompatible with the photonic structuring required
for efficient extraction. The planar metallo-dielectric antenna circumvents this
trade-off because all structuring is confined to the substrate, leaving the host
crystal itself unprocessed. Realising near-unity collection in practice
nonetheless places two stringent demands on fabrication. First, the SIL must be
bonded to the patterned substrate without a gap, so that the depth of the
nanochannel defines the thickness of the organic layer. Second, the crystal
growth must be controlled precisely to produce large, homogeneous domains with
minimal scattering losses. Only when both conditions are met does the antenna
reach its theoretical collection efficiency~\cite{luo2026designing}.

A further outcome concerns the emitter itself. From the saturated count rate and
the independently measured excited-state lifetime, we estimate a quantum
efficiency of about $90\,\%$ for the investigated molecule, which is, to our
knowledge, the first direct cryogenic estimate for DBT obtained via saturated
excitation, without requiring assumptions on the dipole orientation. Because any
unaccounted collection loss would lower rather than raise this value, it
represents a lower bound, indicating that DBT is close to an ideal emitter in
terms of radiative efficiency. The high collection efficiency translates this
quantum efficiency into the rates reported above and, under pulsed excitation,
into a per-pulse collection probability of approximately $91\,\%$ over the
full spectrum.


An important outcome of our crystal-growth strategy is the strong suppression of fluorescence background. In contrast to earlier protocols~\cite{zirkelbach2022high}, we omit a second melting cycle, which we infer reduces the accumulation of molecules at the crystal interface and thereby suppresses broadband background emission. This interpretation is supported by the measured second-order correlations: under both CW and pulsed excitation, the background-corrected analysis yields a single-photon purity above $97.5\,\%$ even under strong excitation. Because photons from APD breakdown flashes become non-negligible in our near-infrared spectral window, they can artificially raise the central coincidence peak; these values are therefore conservative lower bounds, and the intrinsic purity is expected to lie even closer to unity.

We also benchmark the coherence of the source through HOM interference of
consecutively emitted photons. Under CW excitation, the zero-delay HOM
visibility exceeds $90\,\%$ in the low-saturation regime after correcting for
the detector timing response, and decreases with increasing excitation power,
in line with the onset of photo-induced spectral diffusion. A more direct
estimate is obtained under pulsed excitation, where the interferometer delay is
matched to the pulse separation and the visibility is extracted from integrated
coincidence peaks. After correction for the residual multi-photon admixture, we
obtain an indistinguishability of $88.5\,\%$. A model that additionally accounts
for polarisation imperfections and interferometer imbalance removes these setup
contributions and yields an emitter-limited value of $91.2\,\%$, together with
an excess dephasing rate of $5.7\,\mathrm{\mu s}^{-1}$ that we identify as the
dominant limitation to unity indistinguishability. This is consistent with the
small residual broadening of the zero-phonon line beyond its Fourier limit
observed in spectroscopy.

Taken together, these results establish antenna-integrated DBT molecules as a competitive platform for single-photon generation, combining high brightness, near-unity purity and strong indistinguishability in a broadband and fabrication-tolerant architecture. Because the antenna does not rely on narrowband cavity enhancement, it does not require emitter-specific spectral preselection and should be transferable to a broad range of molecular guest--host systems. This opens a route towards bright molecular sources operating across different spectral windows. More broadly, the present architecture provides a practical foundation for scaling organic quantum photonics beyond isolated emitters. In particular, multiple molecules within a sub-wavelength distance could be tuned into resonance, for example, by Stark tuning \cite{trebbia2022tailoring,rattenbacher2023chip} or controlled laser-induced frequency tuning \cite{colautti2020laser,lange2024superradiant}, enabling the generation of more complex photonic states while preserving efficient collection. We therefore expect this approach to provide a versatile interface between organic emitters and future quantum photonic networks.

\section{Methods} \label{methods}

\subsection{Preparation of DBT:\textit{p}-DCB} \label{subsec_dbt_pdcb}

DBT-doped \ce{\textit{p}-DCB} crystals are prepared by a melt--quench procedure. A vial containing a mixture of \ce{\textit{p}-DCB} and DBT powders is placed inside a metal cylinder and heated uniformly on a hot plate to $70$--$80\,^\circ\mathrm{C}$, well above the melting point of \ce{\textit{p}-DCB} ($54\,^\circ\mathrm{C}$). During heating, the vial is removed briefly at regular intervals and gently shaken to ensure homogeneous mixing of the dopant molecules. Once the material is fully molten, the vial is transferred into a beaker of cooling water to rapidly solidify the mixture. The resulting solid is then ground into a fine powder.

Using the same procedure, a series of diluted mixtures is prepared by adding further \ce{\textit{p}-DCB} to the stock material. A final DBT concentration of $0.02\,\text{ppm}$ is chosen, which provides sufficient spectral and spatial isolation of individual molecules under vibronic excitation.

\subsection{Fabrication of the antenna substrate} \label{subsec_antenna_substrate}

The antenna substrates are fabricated in two lithography steps \cite{luo2026designing}. In the first step, the outer region of the structure is defined by direct-write laser lithography, followed by dry etching with \ce{CHF3} to form a channel of uniform depth of $420\,\text{nm}$. In the second step, a reduced pattern is aligned to the pre-etched channel area, defining the reflective central region shown in Fig.~\ref{fig_introduction}b. After development, the substrate undergoes a descum step to remove residual resist before sputter deposition.

Three layers are then deposited sequentially: a thin \ce{Cr} layer serving as an adhesion layer, followed by \ce{Ag} and \ce{SiO2}, which form two functional parts in the antenna. Because the channel thickness is critical to the antenna performance, the substrates are inspected by profilometry after lift-off.

\subsection{Direct bonding} \label{subsec_bonding}

Direct bonding through intermolecular interactions provides an effective route to forming a monolithic structure, thereby preventing infiltration of the molten crystal into the fragile interface and ensuring a well-defined organic layer thickness. Before bonding, the SIL and the substrate are activated with mild \ce{O2} plasma. The resulting silanol-terminated surfaces are then brought into contact and pre-bonded at room temperature under externally applied pressure. To further strengthen the bonding, a post-bonding annealing step is carried out at elevated temperature. In our case, the bonded assembly is heated under vacuum at a constant temperature of $150\,^\circ\mathrm{C}$ for several hours. This treatment increases the interfacial bond strength and ensures mechanical and chemical stability against external forces and solvent exposure.

\subsection{Growth of the crystal layer} \label{subsec_crystal_growth}

Crystal growth inside the channel is controlled through a carefully tuned melting and cooling sequence. Separate voltages are applied to the two ends of an indium tin oxide (ITO)-coated glass slide, generating an adjustable temperature gradient along the slide by Joule heating. Initially, identical voltages are applied to both sides, producing uniform heating across the sample. This first heating stage begins before the antenna is placed on the heating plate and continues until the crystal is fully molten. Both voltages are then reduced to values below the melting point in order to avoid unintentional remelting. Subsequently, the voltage on one side is lowered further to establish directional crystal growth from the colder side towards the hotter side, resulting in elongated crystalline stripes.

Because \ce{\textit{p}-DCB} is the only birefringent material in the structure, the resulting crystal domains can be visualized in a cross-polarized microscope image, where they appear bright, as shown in the lower inset of Fig.~\ref{fig_introduction}a.

\subsection{Optical characterization of DBT molecules} \label{subsec_characterization}

The antenna sample containing DBT:\ce{\textit{p}-DCB} is cooled to approximately $1.4\,\mathrm{K}$ in a helium cryostat (UHV bath cryostat, CryoVac). In the excitation path, the laser beam reaches the sample through a stack of four wedged windows and is tightly focused onto the emitter by an aspheric lens (355330, LightPath, $\mathrm{NA}=0.77$). The selected molecule is brought into the laser focus using three cryogenic-compatible nanopositioners (Attocube) for coarse positioning, followed by fine alignment with a fast steering mirror.

\subsection{Brightness measurements} \label{subsec_brightness}

The $0$-$0$ ZPL fluorescence is isolated using a tunable notch filter (BNF-746-OD4-12.5M, OptiGrate). The filter is operated at an incidence angle of approximately $6^\circ$, for which the reflectivity exceeds $90\,\%$ at the 0-0 ZPL frequency of the target molecule. The filtered signal is then directed to an APD (Excelitas) with an active area of $(180 \times 180)\,\mathrm{\mu m}^2$ and a measured detection efficiency of $50.6\,\%$, as calibrated using an attenuated laser. A larger active area was deliberately chosen to accommodate the expanded focal spot caused by chromatic aberration when detecting broadband fluorescence. This trade-off between detector size and detection efficiency ensures a more reliable determination of both the photon collection rate and the quantum efficiency of the DBT molecules.
\subsection{Photon autocorrelation measurements} \label{subsec_g2}

Photon autocorrelation measurements are performed in an HBT configuration, in which the incoming photon stream is split by a pellicle beam splitter. The use of a pellicle minimizes parasitic reflections associated with thicker beam-splitting optics. Photons in the two output arms are detected by APDs (Laser Components) with detection efficiencies of $65\,\%$ at the $0$-$0$ ZPL wavelength. Their combined two-detector IRF has a width of approximately $2\,\text{ns}$. The APDs used in the HBT and HOM measurements have nearly identical characteristics and differ mainly in their optical coupling geometry, namely free-space versus fibre-coupled operation. This close similarity allows us to use the independently measured second-order correlation function to correct and model the HOM data during curve fitting.

\subsection{Photon indistinguishability measurements} \label{subsec_hom}

Photon indistinguishability is measured using a fibre-based HOM interferometer. The photons are first prepared in diagonal polarization and coupled into the setup through a single-mode fibre. They then pass through a fibre polarizing beam splitter, which balances the intensity between the two arms of the Mach--Zehnder interferometer. Parallel and orthogonal polarization configurations are generated using a three-paddle fibre polarization controller optimized by gradient descent. One interferometer arm additionally contains a fibre delay of approximately $171\,\text{ns}$, allowing photons emitted in consecutive excitation cycles to interfere at the second beam splitter. The two output ports of this beam splitter are connected to two fibre-coupled APDs (Laser Components) for correlation measurements.

\backmatter

\bmhead{Acknowledgements}

The authors acknowledge financial support from the Max Planck Society, the Deutsche Forschungsgemeinschaft (DFG, German Research Foundation) -- ID 429529648 -- TRR 306 QuCoLiMa (“Quantum Cooperativity of Light and Matter”) and the Free State of Bavaria via the Munich Quantum Valley lighthouse project ”QuMeCo”. We thank Vahid Sandoghdar for valuable discussions and continuous support. T.H. is part of the Max Planck School of Photonics supported by the German Federal Ministry of Research, Technology, and Space (BMFTR), the Max Planck Society and the Fraunhofer Society. The authors acknowledge the use of Claude Opus 5 to assist with language editing, grammar correction, and improving overall readability of the manuscript. The authors reviewed and edited all content generated by this tool and take full responsibility for the accuracy, data integrity, and final text of the published work.

\begin{itemize}

\item\textbf{Author contributions.} S.G.\ conceived the study. S.L.\ and J.R.\ designed and fabricated the antennas with assistance from T.H. S.L.\ established the cryogenic single-molecule setup. T.H. conceived and implemented the pulsed-excitation and two-photon-interference experiments, and developed the analysis correcting the correlation and interference data for the detector timing response. Experiments were performed and analyzed by S.L.\ and T.H.\ under the guidance of T.U., A.S., and S.G.. S.L., T.H.\ and S.G.\ wrote the manuscript with input from all authors.
 \end{itemize}

\bibliography{sn-bibliography_2} 

\end{document}